# SYNTHESE DES CONTROLEURS OPTIMAUX POUR LES SYSTEMES A EVENEMENTS DISCRETS.

**Prof. Jean Marie MOANDA et Ir. Junior BAKOLA MONGO**
Département de Génie Electrique et Génie Informatique
Université de Kinshasa, Faculté Polytechnique

**RESUME**

Dans cet article, on introduit le problème de synthèse de contrôleurs optimaux des systèmes à événements discrets et nous proposons une procédure pour la résolution de ce problème, dans le cas où le procédé et les spécifications sont représentés par des automates à états finis et présentant une complexité croissante.

Nous allons souscrire à la méthodologie de synthèse selon la théorie de la commande par supervision initiée par Ramadge et Wonham. Par une illustration sur un exemple simple, puis sur un modèle offrant une complexité élevée.

Dans cet esprit, les langages, méthodes et outils de développement mis en œuvre pour les spécifier et les développer doivent s'élever à un niveau de qualité permettant de satisfaire les exigences exprimées. Face à cette situation, nous contribuons dans ce travail à systématiser l'emploi des méthodes formelles dans les cycles de développement des systèmes en l'outillant et l'adaptant au langage UML (Unified Modeling Language), qui est le plus exploité dans les projets industriels.

**MOTS CLES :** Systèmes à Evénements Discrets, Synthèse des contrôleurs optimaux

## INTRODUCTION

Depuis quelques décennies les progrès technologiques ainsi que les exigences d'une compétitivité sans cesse croissante ont présidé à l'apparition de systèmes de plus en plus complexes ainsi là où par le passé le bon sens suffisait, il est devenu de plus en plus crucial de disposer d'outils formels qui permettent d'analyser de tels systèmes c'est notamment le cas des systèmes de production dont la complexité rend parfois difficiles leur dimensionnement, leur automatisation et leur gestion.

Certains s'attachent aux aspects de sureté de fonctionnement d'un système automatisé tel que la fiabilité, la défaillance mais vu les progrès technologiques et les concurrence qui ne cesse de croitre sur le marché, un nouvel aspect d'analyse et de commande est née celui de la synthèse de contrôleur.

La méthodologie s'intéresse au fonctionnement normal et désiré du système, elle a été introduite par Ramadge et Wonham.

Notre travail sera alors porté sur l'application de la méthode de synthèse de contrôleur sur le SED, en usant des méthodes formels couplés aux techniques de génie logiciel pour ainsi débouché sur un modèle offrant clarté et concision du point de vue analyse et conception.

## PRELIMINAIRES

Nous nous situons dans le contexte de contrôle par supervision, défini par la théorie de Ramadge et Wonham. Etant donnés un procédé P et une spécification de fonctionnement Sspec, on souhaite synthétiser un contrôleur C de façon à ce que le système en boucle fermée C/P, respecte la spécification. C'est-à-dire qu'on doit chercher le langage $L(P) \cap L(Sspec)$.

Ce langage appelé fonctionnement désiré correspond à l'ensemble des séquences qui peuvent être générées par le procédé et qui sont tolérées par la spécification, ce langage est noté $L_D$.

Il n'est pas toujours possible (prise en compte d'événements incontrôlables $\Sigma_u$) de restreindre, par le contrôle, le fonctionnement d'un procédé à n'importe quel sous langage de ce fonctionnement.

L'existence d'un contrôleur C tel que $L(C/P) = L_D$ réside dans le concept de contrôlabilité [WON 87].

A partir des modèles accepteurs P d'un procédé et $S_{spec}$ d'une spécification de fonctionnement, l'algorithme de KUMAR permet de vérifier la contrôlabilité du langage de spécification $L(S_{spec})$. De plus, dans le cas où le langage $L(S_{spec})$ n'est pas contrôlable, cet algorithme permet de synthétiser un modèle accepteur du langage suprême contrôlable du fonctionnement désiré, c'est-à-dire $SupC(L_D)$ [Kumar et al. 91].

## PROBLEMATIQUE DU CONTROLE

Le problème de la synthèse de contrôleur peut se résumer ainsi :

Etant donné un système, qui modélise un programme ou un système réel comment forcer ce système à respecter ses spécifications, en restreignant son comportement le moins possible ?

La première question qui se pose est l'implémentation du système contrôlé.

La vision adoptée est celle du raffinement d'une spécification incomplète, afin de la rendre correcte vis-à-vis d'exigences bien définies. Il s'agit typiquement de modifier les transitions du système initial ou bien contraindre le comportement à l'aide d'un contrôleur qui a des moyens d'action et d'observations sur le système. Il s'avère en fait que les problèmes sont souvent équivalents, mais il convient de spécifier le point de vue, adopté.

**CONTRIBUTION**

Soit un cahier de charge, la synthèse d'un contrôleur optimal pour un tel système n'est sans doute pas difficile en recourant à la théorie **RW**. Il s'avère alors difficile si le système traité est un système complexe présentant un problème d'explosion combinatoire. Notre modeste contribution consiste alors à user d'un certains nombres des diagrammes UML, pouvant rendre le problème complexe facilement modélisable puis finalement obtenir un modèle FSM conforme au un langage formel.

Nous avons choisi pour cela le modèle en peau d'oignon qui est maintenant normalisés au sein d'UML. Ils permettent de combiner des automates à états finis suivant deux principes: le "parallélisme" (automates concurrents) et l'inclusion hiérarchique. L'inclusion signifie que chaque état peut être décomposé en un automate qui décrit le comportement avec une granularité plus fine. Les transitions sont étiquetées avec des événements déclencheurs, des actions (qui sont des événements déclenchés) et des conditions.

L'utilisation d'UML dans l'industrie recouvre principalement les diagrammes de cas d'utilisation, les diagrammes de classe, les diagrammes de séquence et les diagrammes d'activité.
Dans notre contexte, nous utilisons les diagrammes d'activités, car nous nous intéressons principalement aux comportements des systèmes modélisés. Les diagrammes d'activité permettent de modéliser des flux de contrôle : signaux, données, algorithmes ou procédures.
Les diagrammes de comportement, qui décrivent les aspects dynamiques d'un système ou processus métier. Ils comprennent principalement les diagrammes d'activité, d'états-transitions et de cas d'étude.

Nous nous proposons d'appliquer les méthodes formelles, selon deux approches complémentaires.
La première consiste à appliquer des techniques au cours de sa phase d'analyse, pendant laquelle les exigences de fonctionnement sont formalisées. Cette démarche intervient en amont de la phase de conception détaillée du système.
Il s'agit de déterminer, à partir d'un ensemble de configurations initiales du système, une stratégie permettant de maintenir son fonctionnement dans l'ensemble de ses comportements admissibles, quelle que soit l'influence de son environnement. L'objectif est d'abord de comprendre la dynamique du comportement du système à réaliser, afin d'aider l'ingénieur à concevoir ensuite un modèle précis de ce système que nous pourrions vérifier.
La deuxième approche concerne la vérification formelle de propriétés attendues sur le modèle du système, obtenu en phase de conception détaillée.

Notons que les deux approches permettent d'obtenir le diagramme d'états du contrôleur et suivant l'algorithme de KUMAR on arrive à un contrôleur optimal qui garantisse les respects des spécifications. Ce dernier permettra d'obtenir une implémentation et une simulation aisée sur l'environnement MATLAB.

**IMPLEMENTATION AVEC MATLAB – SIMULINK**

Simulink est un logiciel qui permet de modéliser, simuler et analyser des systèmes dynamiques (système dont les sorties et les états évoluent au cours du temps).

Dans ce paragraphe, nous allons présenter le nouveau simulateur systèmes qui va interagir avec SIMULIK.
Stateflow est un outil graphique puissant de conception, de développement et pour la modélisation de systèmes de contrôle complexe. Il nous offre la possibilité de modéliser et simuler le comportement de systèmes fondés sur le principe des automates à états finis. Cette particularité de Stateflow va nous permettre de combiner plusieurs éléments d'un système en une simulation de boucle fermée (« closed-loop simulation »).
La combinaison de Stateflow et de Simulink nous offre donc une méthode de simulation à la fois précise et complexe. Stateflow fait partie de Simulink, il permet d'y ajouter des blocs contenant des graphes à nombre d'états fini, avec des étapes, des actions, des transitions, mais également avec des évènements. La simulation du schéma- bloc utilise Visual C++, c'est à dire que les modèles une fois définis sont traduits en C++ et compilés, ce qui donne l'exécutable utilisé pour la simulation. La particularité de deux approches présentées ci-haut est qu'elles nous offrent directement un diagramme d'états du procédé contrôlé facile (c'est-à-dire le produit synchrone des automates du système ascenseur pris individuellement et l'application de l'algorithme de KUMAR est intégré dans le modèle proposé) à construire sous stateflow (états, transitions et événements).

**IMPLEMENTATION DU PROCEDE-CONTROLE**

Les objectifs du contrôle en boucle fermée sont de maintenir le procédé à respecter les spécifications en agissant sur l'occurrence des événements contrôlables.

Nous allons utilisé une autre approche visant à concevoir un système entraîné par les événements qui alors modélise le comportement du système à étudier, qui se trouve être décrit en termes de transitions d'état.
La déclaration des états active est basée sur l'occurrence des événements sous certaines conditions.

Le procédé contrôlé est implémenté de façon modulaire tel que chaque contrôleur ascenseur est programmé graphiquement ensuite les deux supervisé par un contrôleur global qui gère les conflits.

La programmation graphique de chaque contrôleur est réalisée de façon à traduire le diagramme d'état donné en un FSM compatible Stateflow, qui intègre deux importantes tables (Condition+action) permettant la coordination des propriétés d'accessibilité et de co-accessibilité.

La loi de contrôle est codé sous Matlab est imprime son action sur le modèle graphique et gère les occurrences des événements contrôlables et incontrôlables.

**EXEMPLE**

L'exemple retenu pour illustrer notre contribution est un contrôleur du système ascenseur. L'environnement doit contrôler un ensemble de 2 ascenseurs automatiques dans un bâtiment de 6 étages. Nous remarquerons le système ne présente pas seulement un problème de complexité combinatoire mais constitue également un système complexe.

Chaque composant du système est modélisé comme un SED, la grande difficulté est due au fait que ces composants aussi nombreux sont astreints à une série des spécifications liées au cahier de charge. Ce qui nous amène à user d'une approche beaucoup plus conciliante permettant la prise en compte des tous ces composants et leurs spécifications sans toute fois arrivée à atteindre une explosion combinatoire.

La première approche consiste à appliquer des techniques au cours de sa phase d'analyse.

|  | Evénement | Réponse du système |
|---|---|---|
| 1 | un passager potentiel appelle l'ascenseur | 1. le bouton d'appel s'éclaire<br>2. sélection d'un ascenseur<br>3. envoi de l'ascenseur à l'étage d'appel |
| 2 | un passager indique l'étage voulu | 1. le bouton de l'étage s'éclaire<br>2. envoi de l'ascenseur à l'étage demandé |
| 3 | un passager met sur Arrêt l'interrupteur Marche/Arrêt | l'ascenseur s'arrête |
| 4 | un passager met sur Marche l'interrupteur Marche/Arrêt | l'ascenseur reprend le traitement des requêtes en cours |
| 5 | un passager obstrue la porte pendant qu'elle se referme | la porte s'ouvre et le timer de fermeture est redémarré |
| 6 | un passager appuie sur le bouton Ouverture de la porte | la porte reste ouverte et le timer de fermeture est remis à 0 |
| 7 | un passager appuie sur le bouton Fermeture de la porte | la procédure de fermeture est démarrée |
| 8 | le timer de fermeture de la porte expire | le cycle de fermeture de la porte débute |
| 9 | un passager appuie sur le bouton d'appel d'urgence | le central de contrôle est notifié |
| 10 | l'ascenseur arrive à l'étage | 1. l'éclairage du bouton s'éteint<br>2. le cycle d'ouverture de la porte débute |
| 12 | 'ascenseur quitte l'étage | l'ascenseur va à l'étage le plus proche dans sa liste de destinations |

Tableau1 : Evénements prise en compte dans l'analyse du modèle

Identification de cas d'utilisation et écriture de cas d'utilisation.

| étape | Message | Action | passager 1 | passager 2 | ascenseur |
|---|---|---|---|---|---|
| 0 | l'ascenseur est inactif au 1er | | | | |
| 1 | demande de l'ascenseur au 4ème pour monter | l'ascenseur démarre vers le 4ème | source | | cible |
| 2 | l'ascenseur passe au 2ème étage | | | | |
| 3 | demande de l'ascenseur au 2ème pour descendre | mise en attente de la requête | | source | cible |
| 4 | l'ascenseur arrive au 4ème | | cible | | source |
| 5 | le passager 1 entre dans l'ascenseur et demande le 6ème étage | l'ascenseur ferme la porte et démarre vers le 6ème | | | |
| 6 | l'ascenseur arrive au 6ème et ouvre la porte | | cible | | source |
| 7 | le passager 1 sort | l'ascenseur ferme la porte et démarre vers le 2ème | | | |
| 8 | le timer de fermeture de la porte expire | | | | |
| 9 | l'ascenseur arrive au 2ème et ouvre la porte | | | cible | source |
| 10 | le passager 2 entre dans l'ascenseur et demande le 1er étage | l'ascenseur ferme la porte et démarre vers le 1er | | source | cible |
| 11 | l'ascenseur arrive au 1er et ouvre la porte | le passager 2 sort | | cible | source |
| 12 | l'ascenseur ferme la porte et se met en état inactif | | | | |

Tableau 2: Différents cas d'utilisation identifiés

Diagramme de séquence

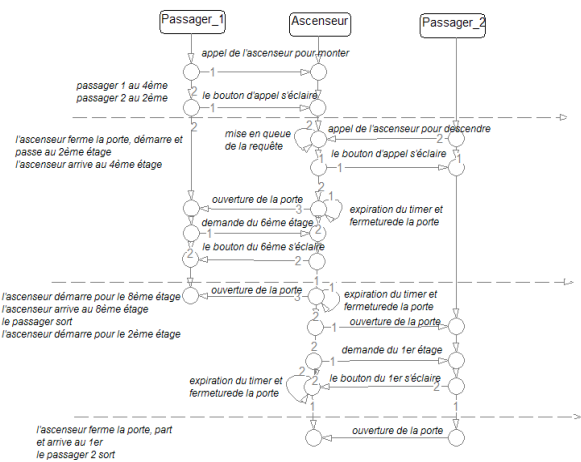

Figure 1 : Diagramme de séquence

Diagramme de cas d'utilisation

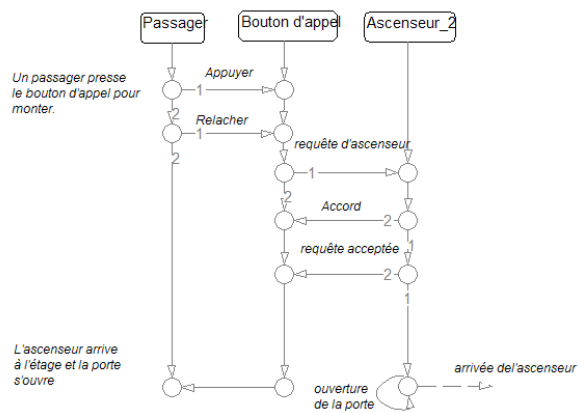

Figure 2 : scénario provenant du cas d'utilisation

## Identification des associations entre objets

| Source des messages | Destination des messages | Message |
|---|---|---|
| bouton d'appel | contrôleur | demande d'ascenseur |
| contrôleur | ascenseur | demande de status |
| contrôleur | ascenseur | ajout de la destination |
| ascenseur | contrôleur | accord |
| ascenseur | Indicateur d'arrivée d'ascenseur | signal d'arrivée |
| capteur de palier | ascenseur | localisation |
| capteur de tension de câble | bloqueurs | engager |
| station de contrôle | bloqueurs | relâcher |
| capteur de tension de câble | station de contrôle | alarme |
| ascenseur | station de contrôle | status |
| bouton de demande d'étage | ascenseur | ajout de la destination |
| interrupteur marche/arrêt | ascenseur | marche/arrêt |
| interrupteur marche/arrêt | station de contrôle | marche/arrêt |
| bouton d'alarme | station de contrôle | alarme |
| ascenseur | porte | ouvrir/fermer |

Tableau 3. Identification et association entre objets

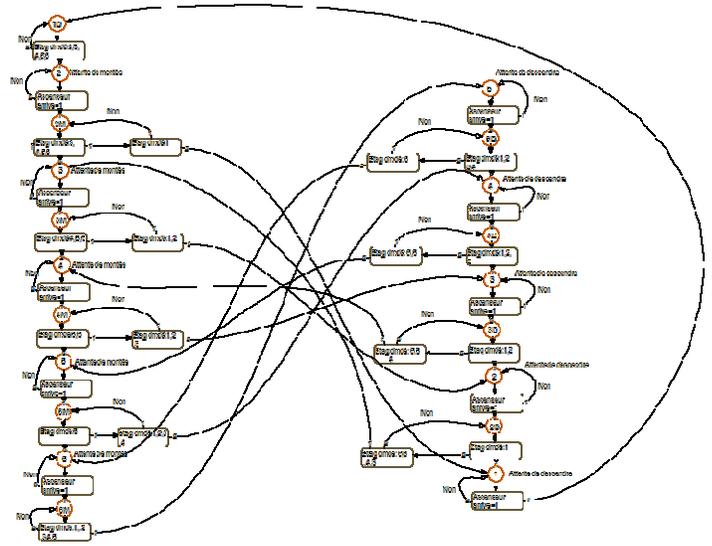

Figure 3 : Automate d'un contrôleur local optimisé du modèle ascenseur

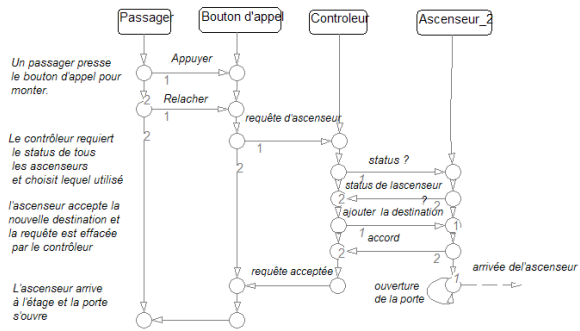

Figure 4 : contrôleur d'un scénario provenant du cas

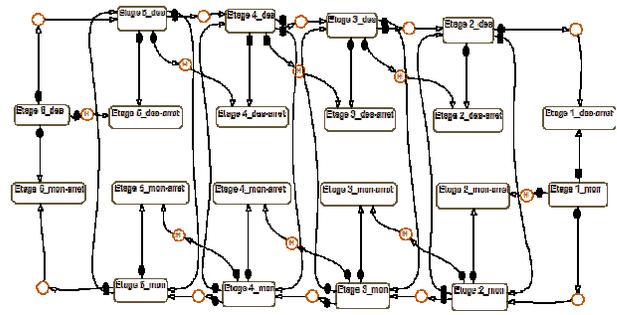

Figure 5 : Automate à état fini de l'ascenseur

## Implémentation sous MATLAB

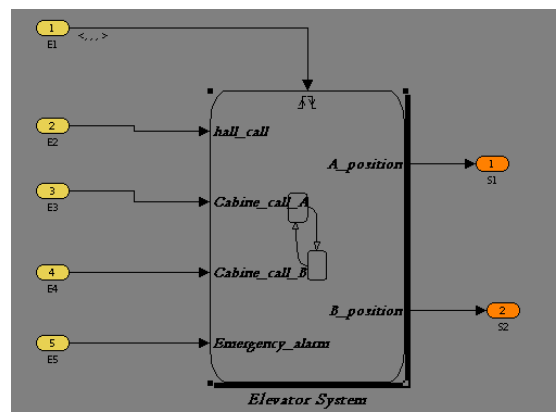

La phase de conception détaillée.

Une fois l'analyse est bien finie nous procédons à la conception d'un contrôleur.
Ce diagramme permet de visualiser des automates d'états finis de l'ascenseur, du point de vue des états et des transitions voir figure 4.

Figure 6 : (Modèle stateflow du contrôleur)

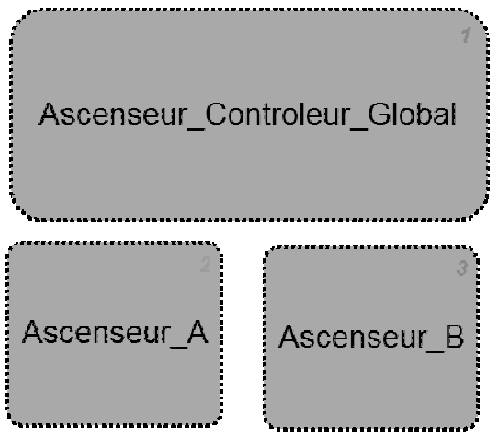

Figure 7 : (Structure modulaire du concept contrôleur sous Stateflow)

Ce point se concentre sur la logique de contrôle du système qui est mis en application dans Stateflow, mais les points suivants sont cruciaux à l'interaction entre Simulink et Stateflow :
- ✓ La logique de contrôle surveille les lectures des capteurs comme entrées de données dans Stateflow.
- ✓ La logique déterminée à partir de ces lectures produit une phrase booléenne d'état.

Le diagramme Stateflow ci-dessous met en application la logique de contrôle de chaque contrôleur pris individuellement puis celle du contrôleur global pris en entièreté.

Le groupe de figures ci-dessus décrit la logique de contrôle illustrée par la machine à état fini du contrôleur d'ascenseur en ne considérant qu'un ascenseur pris isolement, cette logique est donc implémentée dans une machine stateflow sous une programmation graphique traduisant le comportement réel d'un contrôleur d'ascenseur dans un environnement dynamique.

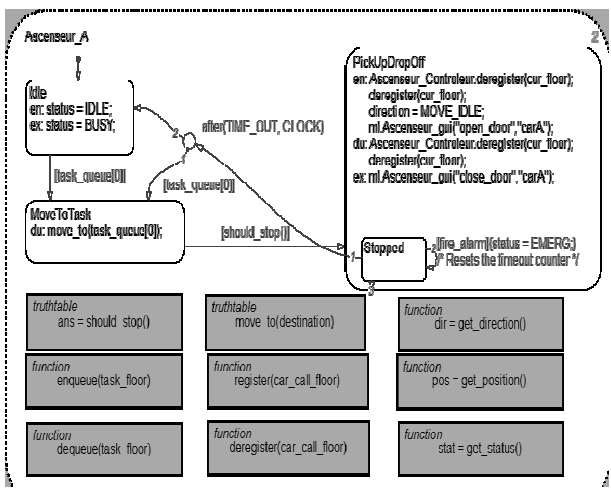

Figure 8 : Le modèle FSM du contrôleur local A d'ascenseur

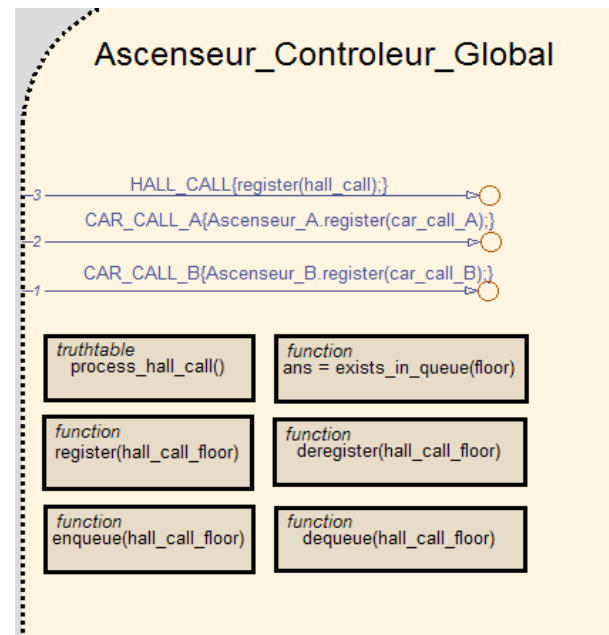

Figure 10 : Logique de commande du contrôleur global du système

Figure 9 : la table de processus du contrôleur local A

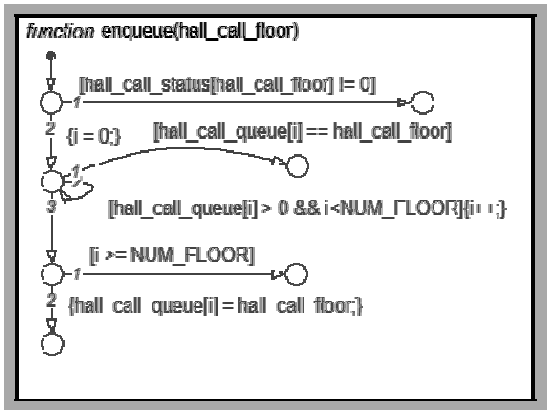

Figure 11 : Le modèle Stateflow d'un FSM de la file d'attente

Si aucun ascenseur n'est disponible, la requête est mise en queue jusqu'à ce qu'un des ascenseurs satisfasse la requête émise. Une fois pressé, le bouton d'appel s'éclaire pour indiquer qu'une requête est en attente. Presser un des boutons d'appel alors qu'une requête pour la même direction est en attente n'aura aucun effet.

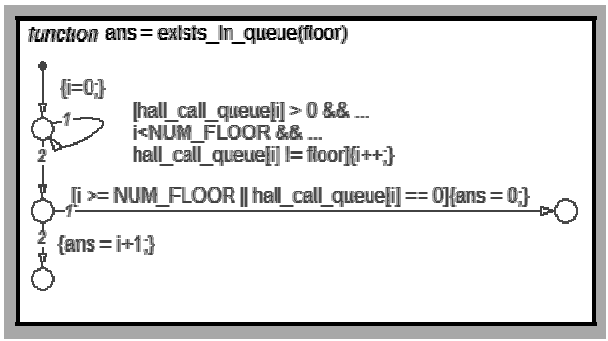

Figure 12 : Le modèle stateflow d'un FSM de statut

Une fois pressé, le bouton d'appel s'éclaire pour indiquer qu'une requête est émise, le contrôleur prend directement en charge la requête et demande le statut d'ascenseur une fois que le statut lui est donné, le système répondra à une requête d'ascenseur en envoyant l'ascenseur le plus proche et qui est inactif ou qui se déplace dans la direction demandée. Si aucun ascenseur n'est disponible, la requête est mise en queue.

## PERSPECTIVES

Comme perspectives nous aurions voulu faire de la synthèse d'appel optimale sur un modèle d'ascenseur (c'est-à-dire : dans le modèle traité, les passagers empruntent la première cabine disponible, quelque soit l'étage auquel ils se rendent, de nombreux arrêts sont nécessaires pour libérer la cabine tandis que avec le contrôle d'appel de destination, les passagers sont regroupés avant de pénétrer dans l'ascenseur. Ceux qui se rendent au même étage parviennent directement à destination, sans arrêt intermédiaire. Les étapes étant moins nombreuses, la cabine est plus rapidement disponible).

## CONCLUSION

La conception d'un contrôleur pour un cas complexe difficilement concevable avec les méthodes classiques développées par INRIA et R&W. Là on a pu recourir à des méthodes développées en génie logiciel enfin d'obtenir un automate du contrôleur optimal et faisant face aux problèmes liés à la compilation de modèle de taille importante nous avons déterminée le modèle du contrôleur qui reflète directement les spécifications choisies.

Ensuite nous nous sommes focalisés sur la conception d'un simulateur pour notre système logiciel intégrant la logique synthétisée dans la seconde partie. La procédure de synthèse présente beaucoup d'avantage car elle nous offre directement le FSM adapté pour être implémenté dans l'environnement utilisé pour concevoir le simulateur. Nous pouvons dire que la formulation de la technique de synthèse génère des contrôleurs corrects, optimaux et offre la possibilité de faire des contrôles différents sur un même modèle (Contrôleurs locaux et global).